\documentclass[letterpaper]{article} 
\usepackage{aaai24}  
\usepackage{times}  
\usepackage{helvet}  
\usepackage{courier}  
\usepackage[hyphens]{url}  
\usepackage{graphicx} 
\usepackage{natbib}  
\usepackage{caption} 
\usepackage{algorithm}
\usepackage{algorithmic}

\usepackage{newfloat}
\usepackage{listings}
\DeclareCaptionStyle{ruled}{labelfont=normalfont,labelsep=colon,strut=off} 
\floatstyle{ruled}
\newfloat{listing}{tb}{lst}{}
\floatname{listing}{Listing}
\usepackage{algorithm}
\usepackage{algorithmic}
\usepackage{natbib}
\usepackage{amssymb}
\usepackage{bm}
\usepackage{amsmath}
\usepackage{wasysym}
\DeclareMathOperator*{\argmax}{arg\,max}

\DeclareMathOperator*{\ie}{i.e.}

\usepackage{wasysym}

\title{Hot or Cold? Adaptive Temperature Sampling for Code Generation with Large Language Models}
\author{
    Yuqi Zhu\textsuperscript{\rm 1},
    Jia Li \male \textsuperscript{\rm 2},
    Ge Li \textsuperscript{\rm 2}\footnote{Corresponding author},
    YunFei Zhao\textsuperscript{\rm 2},
    Jia Li\textsuperscript{\rm 2},
    Zhi Jin\textsuperscript{\rm 2},
    Hong Mei\textsuperscript{\rm 2,3}
}
\affiliations{
    \textsuperscript{\rm 1}Academy of Military Sciences, China\\
    \textsuperscript{\rm 2}Key Laboratory of High Confidence Software Technologies (Peking University), Ministry of Education; School of Computer Science, Peking University, Beijing, China\\
    \textsuperscript{\rm 3}Advanced Institute of Big Data, Beijing, China\\
    zhuyuqi1997@126.com,
    lijia@stu.pku.edu.cn, 
    \{lige, zhaoyunfei, lijiaa, zhijin, meih\}@pku.edu.cn

}

\usepackage{bibentry}

\begin{document}

\maketitle

\begin{abstract}
Recently, Large Language Models (LLMs) have shown impressive abilities in code generation. However, existing LLMs' decoding strategies are designed for Natural Language (NL) generation, overlooking the differences between NL and programming languages (PL). Due to this oversight, a better decoding strategy for code generation remains an open question. In this paper, we conduct the first systematic study to explore a decoding strategy specialized in code generation. With an analysis of loss distributions of code tokens, we find that code tokens can be divided into two categories: challenging tokens that are difficult to predict and confident tokens that can be easily inferred. Among them, the challenging tokens mainly appear at the beginning of a code block. Inspired by the above findings, we propose a simple yet effective method: Adaptive Temperature (AdapT) sampling, which dynamically adjusts the temperature coefficient when decoding different tokens. We apply a larger temperature when sampling for challenging tokens, allowing LLMs to explore diverse choices. We employ a smaller temperature for confident tokens avoiding the influence of tail randomness noises. We apply AdapT sampling to LLMs with different sizes and conduct evaluations on two popular datasets. Results show that AdapT sampling significantly outperforms state-of-the-art decoding strategy. 
\end{abstract}

\section{Introduction}

Code generation aims to automatically generate a program that satisfies a natural language requirement \cite{SkCoder, CodeEditor, ICL}.
In recent years, Large Language Models (LLMs) have attracted great attention for their potential for automating coding \cite{SCoT, AceCoder}.
\begin{figure}[htbp]
    \centering
    \includegraphics[width = 7.8cm]{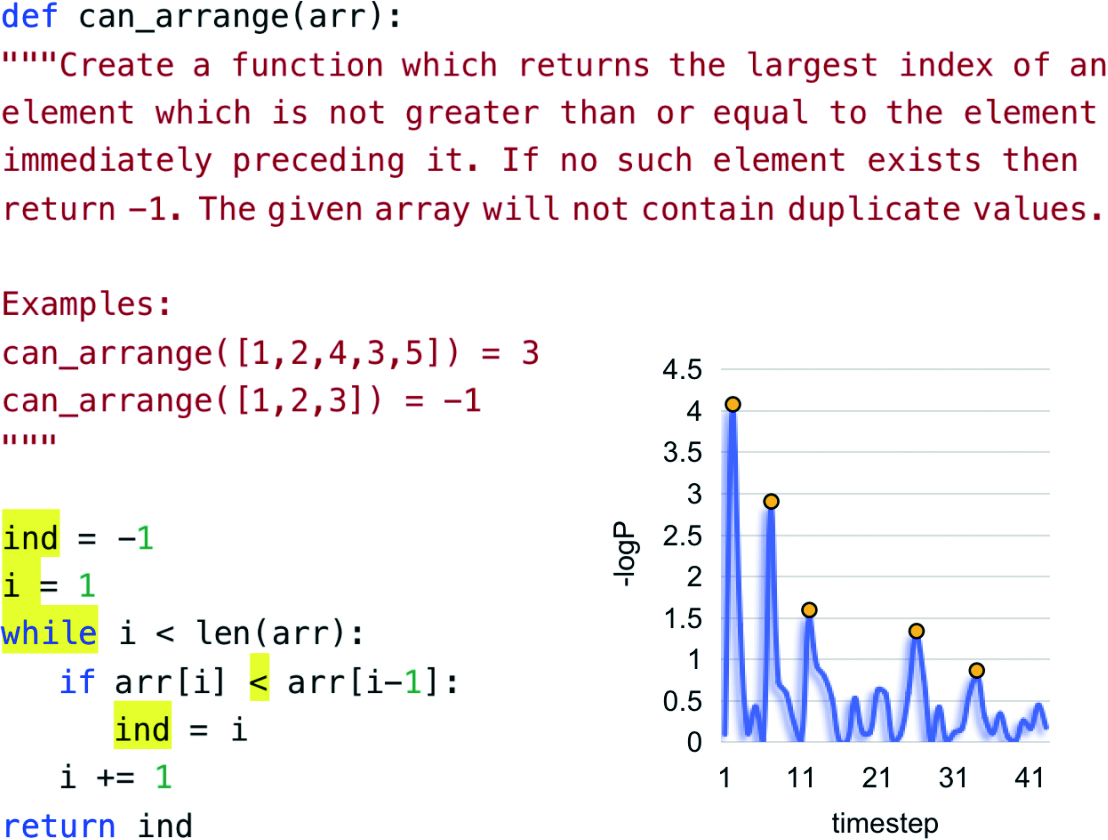}
    \caption{An illustration of a code snippet and its corresponding loss distribution. The challenging tokens are highlighted in yellow.}
    \label{example}
\end{figure} 
Noteworthy models like AlphaCode \cite{AlphaCode} and Codex \cite{codexx} have demonstrated their impressive ability to solve unforeseen programming challenges.

LLMs rely on a decoding strategy to generate code. Existing LLM's decoding strategies for code generation mainly fall into two categories. The first category is search-based methods, which aim to maximize the probability of the next generated token, including greedy search \cite{greedy} and beam search \cite{beamsearch}. Nonetheless, the results generated by these methods lack diversity and are prone to generating empty or repetitive results \cite{inverse}.
The second category is sampling-based methods, which randomly sample the next token based on the probability distribution.
The state-of-the-art (SOTA) approach \cite{codexx} uses temperature sampling that reshapes the probability distribution by introducing a temperature coefficient to control the level of sampling randomness.

Despite the promising results, temperature sampling has limitations in code generation. Since it is initially used to generate Natural Language (NL) with a flexible syntax, its effectiveness decreases when transitioning to code generation, which is a high-accuracy demanding task. 
For instance, CodeGeeX \cite{codegeex}, when utilizing temperature sampling, attains 36.0\% Pass@15 on the HumanEval dataset. 
Thus, it is necessary to explore more advanced decoding strategies to improve the accuracy of LLMs in code generation.

In this paper, we present the first systematic study to explore a decoding strategy for code generation with LLM.
Our contributions can be summarized as follows:

\textbf{(1) By analyzing the loss distribution of code tokens, we categorize code tokens into challenging tokens and confident tokens.}
We design comparative experiments to investigate the differences in loss distributions between source code and NL text.
Our analysis reveals that the source code suffers lesser variation in loss values during generation than NL text. 
Next, we visualize the loss distribution of source code. 
We find an apparent discrepancy in loss values among different code tokens. 
In this study, we refer to these tokens with high loss values as challenging tokens. With statistical analysis, we find that challenging tokens frequently appear at the beginning of a code block. 
An illustration of the challenging tokens is shown in Figure \ref{example}. 
The remaining tokens, characterized by low loss values for LLMs, are referred to as confident tokens.

\textbf{(2) In light of our findings, we propose Adaptive Temperature (AdapT) sampling, which dynamically adjusts the temperature coefficient. }
Compared to standard temperature sampling, AdapT sampling dynamically adjusts the temperature coefficient $T$ according to the type of next-tokens.
Our motivation is that LLMs require exploring diverse choices for challenging tokens.
For confident tokens, we should select tokens with high probabilities.
Specifically, for challenging tokens, AdapT sampling utilizes a high $T$ value to increase sample diversity. AdapT sampling uses a low $T$ value for confident tokens to minimize randomness noise.

\textbf{(3) Experimental results show that AdapT sampling can improve the pass@$\bm{k}$ metric on HumanEval and MBPP datasets with different sizes of LLMs.}
We apply the AdapT sampling to multiple LLMs with various sizes (from 2B to 13B) and conduct evaluations on two representative code generation datasets (HumanEval and MBPP).
Experimental results show AdapT sampling outperforms the SOTA decoding strategy which uses a standard temperature sampling. For example, it surpasses the pass@15 metric over the SOTA method by up to 13.6\% on HumanEval.
We further investigate the robustness of AdapT sampling to different hyperparameter settings and the quality of the code generated by AdapT sampling.
Our code is available at \url{https://github.com/LJ2lijia/AdapT}.

\textbf{(4) Future directions.}
Based on our ﬁndings, we list the current challenges and propose future research directions on developing effective decoding strategies for code generation.

\section{Background}
\subsection{Code Generation with LLMs}

LLMs are transformer-based models that are trained using large corpora of NL text and source code. In recent years, LLMs have achieved impressive results in automatic code generation.
Among LLMs, the GPT family of LLMs from OpenAI is popular and powerful, including GPT-3 (175B parameters) \cite{gpt3},  Codex (175B parameters) \cite{codexx}, etc. Since OpenAI LLMs are closed-source, there have been many attempts to reproduce similar LLMs, such as CodeGen \cite{CodeGen}, CodeGeeX \cite{codegeex}, InCoder \cite{incoder}.

\subsection{Decoding Strategy}
Given a requirement $x$, LLMs rely on a decoding strategy to generate the code auto-regressively.
The decoding strategy determines how LLMs select the next token $y_{t}$ based on the context $y_{<t}, x$. $y_{<t}$ is the token sequence that has been generated. 
There has been a series of works exploring decoding strategies for NL generation, and these methods have been subsequently applied to code generation. They can be classified into two categories: search-based and sampling-based methods. 

\paragraph{Search-based Decoding Stratrgy}  
Greedy search \cite{Mou2015OnEP} is one of the most commonly used decoding strategies. In greedy search, the model selects the next token which maximizes the probability: $y_t = \argmax_y p(y_t|y_{<t}, x)$. 
Despite its simplicity, it may lead to overly conservative results and a lack of diversity \cite{greedy_bad}.
Beam search \cite{beamsearch} is an improved version of greedy search. This algorithm retains top $B$ (beam size) tokens with the highest probability. 
However, it has been found that beam search results in degenerations such as repetitions and empty \cite{topp}.

\begin{figure*}[t]
    \centering
    \includegraphics[width = 15cm]{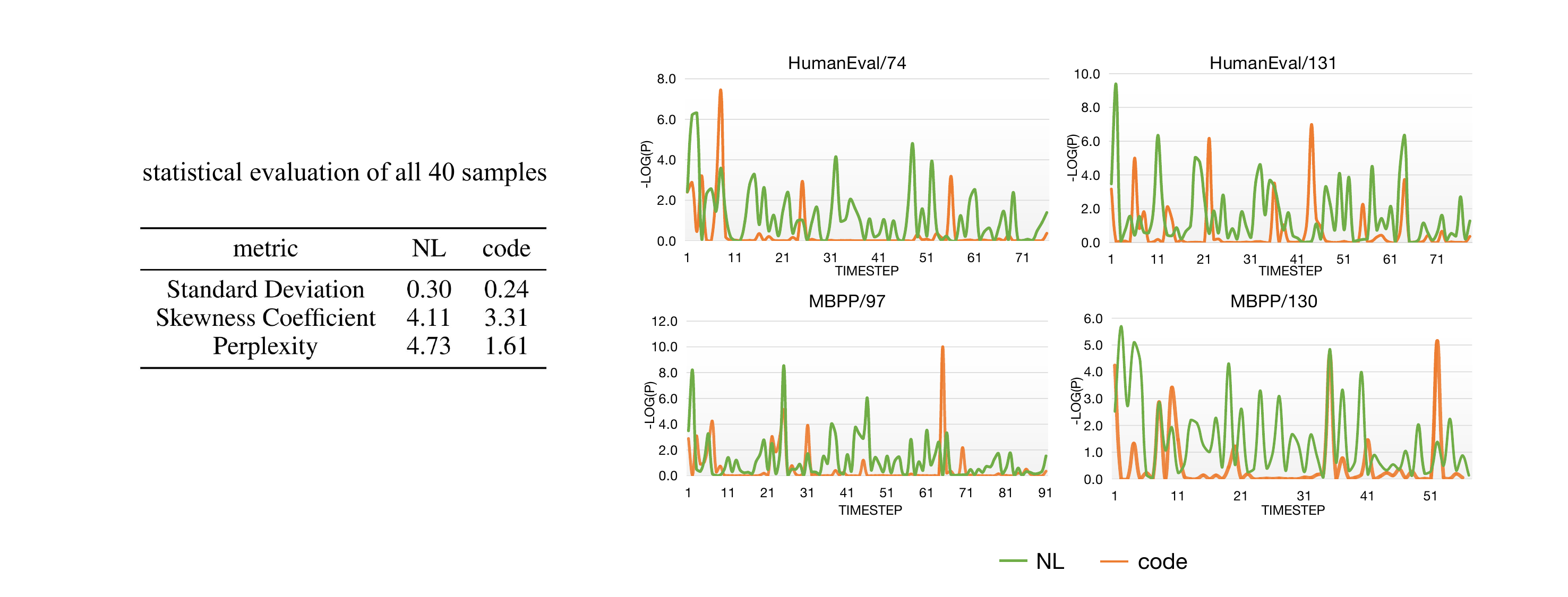}
    \caption{Comparison of loss distributions on NL text and source code. The left table shows the statistical results of different distributions on 40 samples. The right figures show several examples of loss distributions on NL text and source code.}
    \label{p_1}
\end{figure*}
\paragraph{Sampling-based Decoding Strategy}

The degenerations such as empty sequences and repetitions can be alleviated using sampling decoding, which randomly selects the next token based on predicted probability. 

Temperature sampling \cite{temperaturesampling} has been applied widely, it uses a temperature coefficient $T$ (usually $ \in [0,1]$) to control the sampling randomness. 
Given the logits $u$ and temperature $T$, 
the softmax distribution $p'$ is re-estimated as:
\begin{equation}
     p'(y_t|y_{<t}, x) = \frac{\exp(\frac{(u(y_t|y_{<t}, x))}{T})}{\sum_{j=1}^n\exp(\frac{(u(y_j|y_{<t}, x))}{T})}
\end{equation}

In addition, researchers propose Top-$k$ \cite{topk} sampling to further improve performance.   
At each step, Top-$k$ sampling filters the $k$ most probable next tokens and redistributes the probability among these $k$ tokens for sampling. However, the unreliable tail problem in the Top-$k$ sampling may affect the sampling quality. 
Top-$p$ sampling \cite{topp} eliminates the unreliable tail problem by sampling from the smallest token set whose cumulative probability reaches the threshold $p$.

LLMs usually use the method of combining temperature sampling and Top-$p$ sampling to achieve SOTA results \cite{codexx, CodeGen, incoder}. Specifically, the logits are first rescaled with temperature $T$. After this, Top-$p$ sampling is employed to derive the final results.
Existing work \cite{codexx} finds out that temperature coefficient $T$ has an obvious influence on the code generation results.
Increasing the $T$ value can enhance the chance of exploring the correct answers, but this comes at the cost of introducing more errors in the generation results. 
Therefore there is a need to develop more effective sampling methods specifically for code generation.

\section{Analysis of the code generation process}
In this section, we first investigate the differences between NL generation with LLMs and code generation with LLMs. 
We compared NL text's loss distributions (i.e., cross-entropy loss \cite{crossentropy}) to ones of source code. 

Furthermore, we analyze the fluctuations of loss values of code tokens within code snippets. 
We find that code tokens can be categorized into challenging tokens and confident tokens.
Based on the analysis, we discuss the challenges encountered in code generation.

\subsection{Loss Distribution Comparison}
In this section, we analyze the differences between the process of generating NL text and source code with LLMs. We choose loss distribution as the comparison metric.

We conduct experiments with a powerful LLM for source code - CodeGen. We select the CodeGen-mono with 2 billion parameters (CodeGen-2B) as the base model.
This model is trained with a 635GB code corpus and 1159GB English
text data. Therefore, CodeGen can generate NL text and code.
The datasets used in this section are HumanEval  \cite{codexx} and MBPP \cite{MBPP}, which are representative datasets in code generation.

\begin{figure*}[htbp]
\centering
\includegraphics[width=17cm]{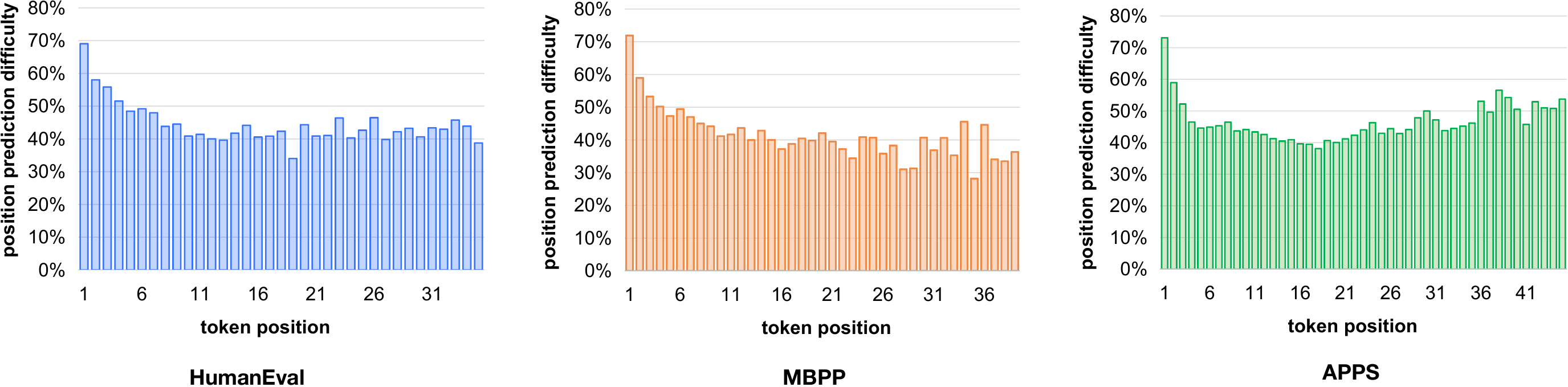}
\caption{The prediction difficulty of different positions, the x-axis represents the position of the token within the line of code, and the y-axis represents the prediction difficulty of the token.}
\label{difficulty_3}
\end{figure*}

FFirst, we randomly select 20 code samples each from HumanEval and MBPP datasets.
Then, we manually write NL descriptions for each code snippet, which describes the functionality of the code. We keep the alignment of the length of NL and code as much as possible while maintaining text fluency. As a result, we obtain 40 NL descriptions aligned with their corresponding code snippets. Next, we gather the loss values of the CodeGen model on these NL descriptions and code snippets, respectively.

We use various metrics (e.g. mean value \cite{mean}, standard deviation, skewness, and perplexity) to compare the loss distributions of NL descriptions and source code.
Standard deviation \cite{standevi} reflects the average amount of variability.
Skewness \cite{skewness} is a measure of the asymmetry of the probability distribution of a real-valued random variable about its mean.
Perplexity \cite{perplexity} is a measurement of how confidently an LLM predicts a sample.
As shown in the table in Figure \ref{p_1}, the average value of losses on NL descriptions is higher than the one on the source code. 
When compared with code, NL suffers greater variation in prediction loss during generation. 
The value of the skewness shows that there are more tokens with large loss values in NL descriptions than in the source code.
The LLM also has a higher perplexity for NL descriptions than for source code. 
We show a few examples in Figure \ref{p_1} to visualize the differences.

These differences arise because source code has a more strict syntax and semantics compared to NL. 
Therefore, when generating code, some tokens can be easily inferred based on grammatical rules, and LLMs can confidently generate these tokens with low loss values. In contrast, NL allows greater freedom in word usage and often presents multiple viable choices for the same context, which results in high loss values.

Additionally, we find that \textbf{a number of peaks occur in loss distributions of source code, i.e., tokens with a higher loss value than nearby tokens. }
A higher loss value means that it is more difficult to make correct predictions at these locations.
We call the tokens with peak loss values \textbf{challenging tokens}.
As for the tokens that have low loss values, the model has more confidence in predicting them correctly. We refer to them as \textbf{confident tokens}.

\subsection{In-depth Study of Code Tokens}
 This section provides a detailed investigation of the challenging tokens and confident tokens.
We analyze samples from the MBPP \cite{MBPP} (500 samples), HumanEval \cite{codexx} (164 samples), and APPS \cite{hendrycks2021apps} (train set: 5000 samples) datasets. We use CodeGen-2B to generate the ground truth code in these datasets and collect the corresponding loss values.

First, we define the predictive difficulty (PD) of a token, which is the rank (\%) of the token loss among all token loss values in the code snippet, and compute PDs for all tokens.
Then, we split each code snippet into code lines and integrate (average) the tokens' PDs in the same position of different code lines to obtain the position prediction difficulty.
Finally, we statistically average the position prediction difficulty across all codes in the dataset. We omit positions where the cumulative data numbers of these positions are below 5\% of the overall token numbers.
Figure \ref{difficulty_3} shows the position prediction difficulty of different token positions. The results reveal the regularity that the tokens in the first position of code lines have the highest prediction difficulty of 69.0\% (HumanEval), 73.1\% (MBPP), and 71.9\% (APPS). 
Therefore, we assume that the first position in each code line is where challenging tokens tend to appear and conduct further experiments.

We distinguish between challenging tokens and confident tokens based on PD.
We introduce a threshold $H$ to separate two types of tokens. Tokens with PD $>H$ are categorized as challenging tokens, while tokens with PD $<H$ are categorized as confident tokens. 
We set $H$=0.9, and the proportion of challenging tokens appearing in the first position was 24.8\% (HumanEval), 22.6\% (MBPP), and 28.1\% (APPS), respectively. 
The results confirm our assumption that the challenging tokens are not randomly distributed for different locations, they tend to appear at the first position of each code line.
We also vary $H$ from 0.5 to 0.9 to investigate the distributions of challenging tokens and confident tokens.
Results are shown in the appendix.

To further investigate the properties of challenging tokens, we distinguish the tokens at the first position of each line based on whether it is the initial token of a code block. 
A code block in Python is a piece of Python program text that can be executed as a unit. 
The code block begins at the first indented statement and continues until the indentation returns to a previous level.
The initial token of the code block has a prediction difficulty of 79.8\% (HumanEval), 83.4\% (MBPP), and 82.5\% (APPS) which is significantly higher than the first positions of other code lines. We derive that among the first positions of code lines, \textbf{the initial token of each code block is the most likely place for a challenging token to appear}. 

This can be attributed to the fact that LLMs need to determine the next control structure after a block of code is presented, which increases prediction difficulty. LLMs can easily generate the next token after receiving the initial token since the strict syntax rules limit the scope of variations. 

\section{AdapT sampling}

In light of our findings, we propose a simple yet effective decoding method, AdapT sampling (Adaptive Temperature Sampling), which adjusts the temperature coefficient $T$ for different tokens. Specifically, for the challenging tokens, which LLMs struggle to predict correctly, 
AdapT sampling uses a high temperature coefficient which introduces more diverse tokens. On the other hand, for confident tokens, the temperature coefficient is set to a small value to minimize randomness noises.
Specifically, we formulate AdapT sampling as follows:
\begin{equation}
    p'(y_t|y_{<t}, x)) = \frac{\exp(\frac{(u(y_t|y_{<t}, x))}{T(t)})}{\sum_{j=1}^n\exp(\frac{(u(y_j|y_{<t}, x))}{T(t)})}
\end{equation}

\begin{equation}
    T(t) = \left\{ \begin{array}{cc}
        a  & \mbox{if \text{$y_t$ is the code block initial token}}\\
        b  & \mbox{else} \\
    \end{array}\right.
\end{equation}
where $T$ is the temperature coefficient and $t$ is the sample timestep. 
$a, b \in [0,1]$ ($a>b$) are hyperparameters that control the degree of sampling randomness. 

\section{Experiments}

\subsection{Benchmarks}

\noindent\textbf{HumanEval} \cite{codexx} is a Python code generation benchmark with 164 test samples. Each sample consists of a manually written programming problem, which consists of a natural language requirement, a function signature, and several unit tests. It asks LLMs to generate the function body based on the requirement and the signature. The unit tests are used to check the correctness of generated functions.

\noindent\textbf{MBPP} \cite{MBPP} contains 500 programming problems collected from real-world communities. Solving these programming problems requires simple numeric manipulations and the basic usage of standard libraries. Each problem contains an English requirement, a Python function signature, and three test cases. We take the requirement and the function signature as input and leverage LLMs to generate the function body. Then, the generated code is evaluated using test cases.

\begin{table}[t]
\centering
  \resizebox{0.85\linewidth}{!}{
  \begin{tabular}{lccccc}
    \hline
    Metric    & pass@5  & pass@10 & pass@15 \\
    \hline
    \textit{CodeGen}: \\
    SP, $T$=0.2  & 29.2 & 31.3 & 32.3 \\
    SP, $T$=0.4  & 33.4 & 37.9 & 40.8 \\
    SP, $T$=0.6  & 33.1 & 38.0 & 40.8 \\
    SP, $T$=0.8   & 32.7 & 39.7 & 43.9 \\
    AdapT       & \textbf{34.4}  & \textbf{40.1} & \textbf{43.9} \\
    \hline
    \textit{InCoder}: \\
    SP, $T$=0.2     & 22.8  & 25.9 & 27.4 \\
    SP, $T$=0.4    & 25.0  & 29.8 & 32.3 \\
    SP, $T$=0.6    & 24.5  & 30.0 & 32.9 \\
    SP, $T$=0.8     & 22.3  & 28.6 & 32.9 \\
    AdapT      & \textbf{25.8}  & \textbf{31.6} & \textbf{35.3}  \\
    \hline
    \textit{CodeGeeX}: \\
    SP, $T$=0.2    & 24.1 & 27.1  & 28.7   \\
    SP, $T$=0.4    & 27.0 & 30.5  & 32.3   \\
    SP, $T$=0.6    & 27.7  &  32.5 &35.4  \\
    SP, $T$=0.8    & 27.1  & 33.0 &36.0 \\
    AdapT   & \textbf{29.4}&  \textbf{36.3}  &  \textbf{40.9}\\
    \hline
  \end{tabular}
     }
\caption{The performance (pass@5, 10, 15) for CodeGen-2B, InCoder-6B, and CodeGeeX-13B on the HumanEval dataset using AdapT sampling and SOTA (SP) method.}
\label{HumanEval_result}
\end{table}

\begin{table}[htbp]

\centering
    \resizebox{0.85\linewidth}{!}{
  \centering
  \begin{tabular}{lccccc}
    \hline
    Metric        & pass@5  & pass@10 & pass@15 \\
    \hline
    \textit{CodeGen}: \\
    SP, $T$=0.2  & 33.1   & 36.5 &  38.4 \\
    SP, $T$=0.4    & 37.0   & 42.1 & 45.0 \\
    SP, $T$=0.6   &  37.1  & 43.5 & 47.0 \\
    SP, $T$=0.8    & 35.2   & 43.1 & 47.0\\
    AdapT    & \textbf{37.2}   & \textbf{44.4} & \textbf{48.2}  \\

    \hline
    \textit{InCoder}: \\
    SP, $T$=0.2  &  23.0 &  27.4  &29.2    \\
    SP, $T$=0.4 & 26.0  &  29.6  & 32.4  \\
    SP, $T$=0.6  &  24.6 &  30.4  & 34.6  \\
    SP, $T$=0.8   &  23.6 &  30.6  &  34.6\\
    AdapT    &  \textbf{26.8} &  \textbf{32.9}  &  \textbf{36.8}\\
    \hline
    \textit{CodeGeeX}: \\
    SP, $T$=0.2    & 20.7   & 23.3 & 22.4 \\
    SP, $T$=0.4  & 23.7   & 28.4 & 31.0  \\
    SP, $T$=0.6  &  24.8  & 31.2 &  34.8  \\
    SP, $T$=0.8    & 25.4   & 31.2 &  35.6  \\
    AdapT    & \textbf{25.8}  &\textbf{32.2} & \textbf{36.0} \\
    \hline
  \end{tabular}
  }
    
    \caption{The performance (pass@5, 10, 15) for CodeGen-2B, InCoder-6B, and CodeGeeX-13B on the MBPP dataset using AdapT sampling and SOTA (SP) method.}
    \label{mbpp_result}
\end{table}

\subsection{Base Models}
In this paper, we select three representative open-source LLMs as base models: CodeGen, InCoder, and CodeGeeX.

\noindent\textbf{CodeGen} \cite{nijkamp2022conversational} is a family of LLMs pre-trained on a large amount of NL texts and source code. It is trained with a 635GB code corpus and 1159GB English text data. 
In this paper, we select the CodeGen-mono with 2 billion parameters as the base model.

\noindent\textbf{InCoder} \cite{fried2022incoder} is pre-trained with a large corpus of permissively licensed code (216GB). It can perform code generation and code infilling. In this paper, we use a version with 6.7 billion parameters, named InCoder-6B, for code generation.

\noindent\textbf{CodeGeeX} \cite{codegeex} is a multilingual LLM with 13 billion parameters. It is pre-trained on a large corpus of more than 20 programming languages and achieves impressive performance on code generation.

Despite OpenAI models' impressive performance, we can not access the probability distribution with only API calls. Therefore, we ignore these models in this paper.

\subsection{Baselines}

In this paper, we choose the SOTA approach mentioned in the background section as the baseline.  
For the sake of brevity, we refer to the baseline (reshaping distribution with temperature sampling) as SP in the following sections.

\subsection{Implementation Details}
We run all of our experiments on 2 NVIDIA V100 GPUs with 32GB memory. For our experimental datasets, the maximum generated length is 500. All experiments are conducted in a zero-shot setting which means we directly feed the input requirement into LLMs without examples. Then we extract generated programs from the model’s output.
\subsection{Metric}

\paragraph{Pass@$\bm{k}$}
Pass@$k$ \cite{codexx} measures the functional correctness of the generated code by executing test cases. We use the unbiased version of pass@$k$, where $n \ge k$ samples are generated for each problem, and $c \leq n$ is the number of correct samples that pass test cases. Pass@$k$ is calculated as follows:
\begin{equation}
    \text{pass@}k=\mathbb{E}_{\mathrm{Problems}}\left[1-\frac{\left( \begin{array}{c} n-c \\ k \end{array} \right)}{\left( \begin{array}{c} n \\ k \end{array} \right)}\right]
\end{equation}

\begin{figure}[htbp]
    \centering
    \includegraphics[width = 7.3cm]{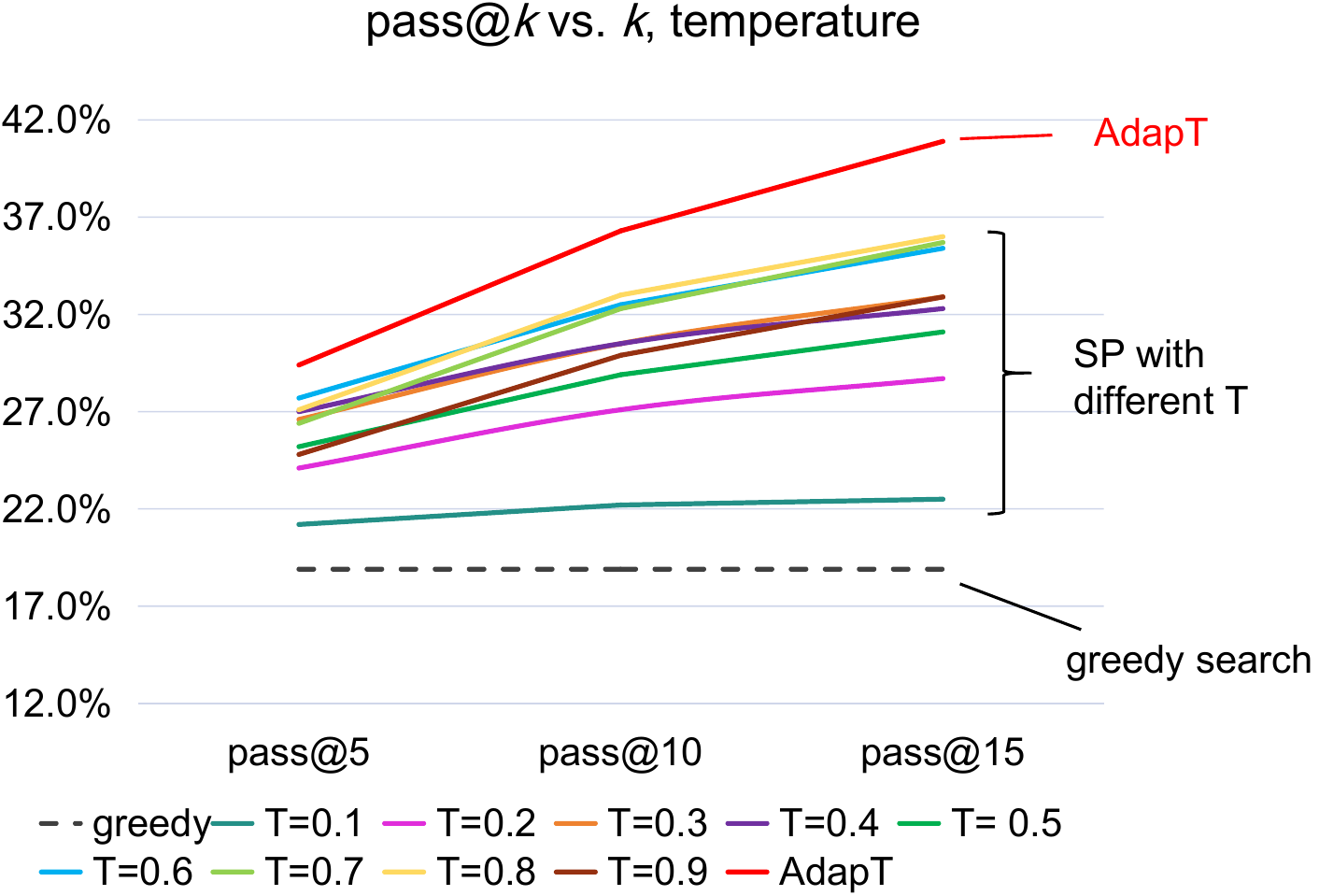}
    \caption{The performance of CodeGeeX-13B on the HumanEval dataset with dense temperature settings.}
    \label{T_0_1}
\end{figure}

\subsection{Main Results}
We apply AdapT sampling to three base models on two code generation datasets. The performance is shown in Table \ref{HumanEval_result} and Table \ref{mbpp_result}. Following the previous works \cite{CodeGen, codegeex, incoder}, we set $p$ of top-$p$ sampling as 0.95 for all our experiments.
Following the previous work \cite{codexx}, we set the values of temperature of baseline as 0.2, 0.4, 0.6, and 0.8 respectively. 
We experimentally select the values of $a$ and $b$, the settings are present in the appendix. The sampling number $n$ is 15. 
The sampling number $n$ is 15. 
As shown in Table \ref{HumanEval_result} and Table \ref{mbpp_result}, AdapT sampling outperforms the SOTA method in terms of pass@5, pass@10, and pass@15 on HumanEval and MBPP datasets. Pass@15 represents the number of problems solved since the sampling number $n$ is 15. 
AdapT sampling with the highest pass@15 indicates that it solves the most problems. 
Notably, on the HumanEval dataset, AdapT sampling can enhance the pass@15 of CodeGeeX from 36.0\% to 40.9\%, reaching a 13.6\% improvement. 
Meanwhile, the number of solved problems increased from 60 to 67. On the HumanEval and MBPP datasets, 
CodeGeeX-13B can solve 14 and 36 previously unsolved problems with AdapT, respectively. 
When run with a constant value of $T$, increasing $T$ will improve the number of problems solved, but it may reduce pass@5 (indicates the proportion of correct answers sampled for each question) due to the introduction of randomness \cite{codexx, topp}. 
On the other hand, AdapT sampling can dynamically adjust the sampling randomness, thus minimizing the noise associated with increasing $T$, hence consistently demonstrating an improvement in pass@5, 10, and 15.

\begin{figure*}[t]
    \centering
    \includegraphics[width = 13cm]{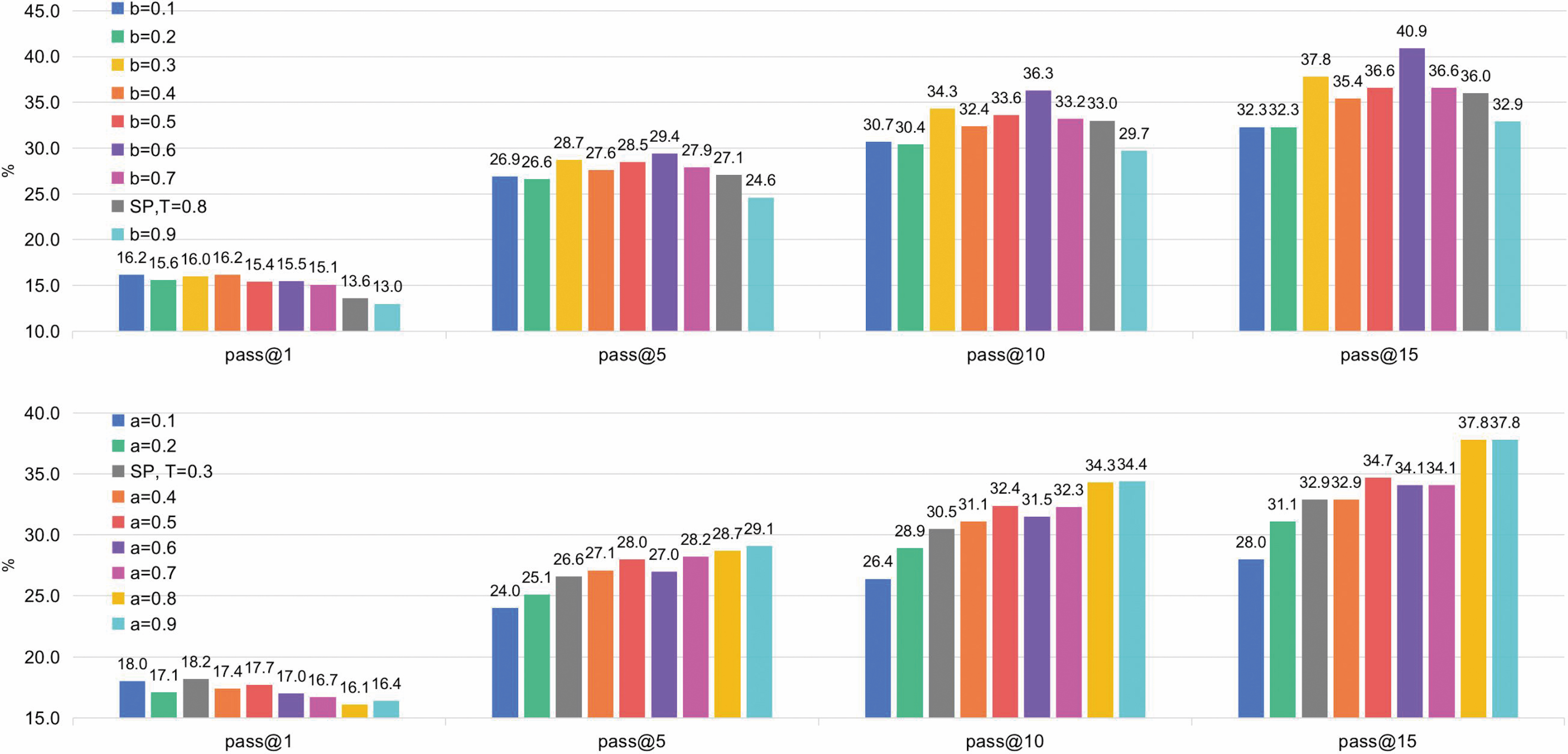}
    \caption{Quantitative analysis of the two hyperparameters ($a$, $b$) in AdapT sampling. The upper half shows the results vary b from 0.1 to 0.9 when fixing $a=0.8$. The bottom half shows the results of fixing b = 0.3 and taking values of $a$ from 0.1 to 0.9. When $a>b$, the AdapT sampling can outperform SP with multiple settings. The model and dataset used in this section are CodeGeeX-13B and HumanEval.}
    \label{hyper}
\end{figure*} 
\subsection{Analysis and Discussion}
We conduct an in-depth and comprehensive analysis of AdapT sampling's capabilities.

\subsubsection{pass@1}

\begin{table}[t]
\centering
  \begin{tabular}{lcc}
    \hline
    Dataset   & HumanEval & MBPP  \\
    \hline
    \textit{CodeGen}: \\
    Greedy Search & 23.8 & 24.0\\
    SP-best  & 22.2 & 24.6\\
    AdapT     &   23.3  & 24.9\\
    \hline
    \textit{InCoder}: \\
    Greedy Search &  14.0 & 12.3\\
    SP-best     & 15.3 & 13.8\\
    AdapT     &  15.2 & 14.8 \\
    \hline
    \textit{CodeGeeX}: \\
    Greedy Search & 18.9 & 13.6 \\   
    SP-best   &18.0 & 13.4 \\
    AdapT   &  18.8 & 13.5 \\
    \hline
  \end{tabular}
     
\caption{Pass@1 results for CodeGen-2B, InCoder-6B, and CodeGeeX-13B using different decoding strategies on HumanEval and MBPP datasets.}
\label{pass1_result}
\end{table}

The pass@1 metric represents the probability that, among multiple pieces of generated code, the first piece chosen would successfully pass a given test case. 
Note that the pass@1 are very strict metrics and are hard to improve. 
We compare the pass@1 results of AdapT sampling with greedy search (which usually has a high pass@1 value) and the best pass@1 results of SOTA (SP-best) and show the results in Table \ref{pass1_result}. Our method outperforms SP-best in 83.3\% cases on the pass@1 metric. Meanwhile, AdapT sampling can reach a comparable pass@1 when compared with greedy search. Greedy search can only sample one answer per question, whereas our method can sample $n$ answers and increase the number of solved questions.

\subsubsection{Compare with Different $T$ Settings}
To explore the upper bound performance of SP, we set the temperature value in SP more densely from 0 to 1, taking a value every 0.1 intervals. The results are shown in Figure \ref{T_0_1}. The performance is not displayed on the resulting graph when $T\ge1$, since it drops significantly when $T\ge1$.
The results of other models can be found in the appendix.
AdapT sampling significantly outperforms temperature sampling at all settings on pass@5, pass@10, and pass@15. 
The LLM can only answer 31 questions correctly with a greedy search. Using the AdapT sampling method, LLM can solve twice as many problems (67) as greedy search.

\subsubsection{Hyperparameters Analysis}
There are two hyperparameters involved in AdapT sampling: $a $ and $b $. In this section, we examine how changing these two parameters affects the code generation results. 
The model and dataset used in this section are CodeGeeX-13B and HumanEval. 

First, we fix $a = 0.8$, and we take a $b$ value every 0.1 step from 0.1 to 0.9, the experimental results are displayed in the upper part of Figure \ref{hyper}. AdapT sampling outperforms SP ($T=0.8$) on pass@1, pass@5, pass@10, and pass@15 metrics when $b$ equals 0.3, 0.5, 0.6, and 0.7. 
Then we set $b=0.3$, and vary the value of $a$ from 0.1 to 0.9 with a step of 0.1. The results are shown at the bottom half of Figure \ref{hyper}. It can be seen from the experimental results that when $a > b$, the results of pass@5, pass@10, and pass@15 can be effectively improved. Additionally, changing a from 0.2 to 0.9 can increase pass@15 from 32.9\% to 37.8\%, achieving a 14.9\% improvement. 

The results indicate that AdapT sampling can outperform SP under a variety of settings, which confirms the robustness of AdapT sampling over hyperparameters.
The empirical hyperparameter guidelines are: for optimizing pass@$k$ ($k>$1), set $a$ to approximately 0.8 and $b$ to around 0.5; for optimizing pass@1, set $a$ to approximately 0.2 and $b$ to around 0.01.

\begin{figure}[t]
    \centering
    \includegraphics[width = 8cm]{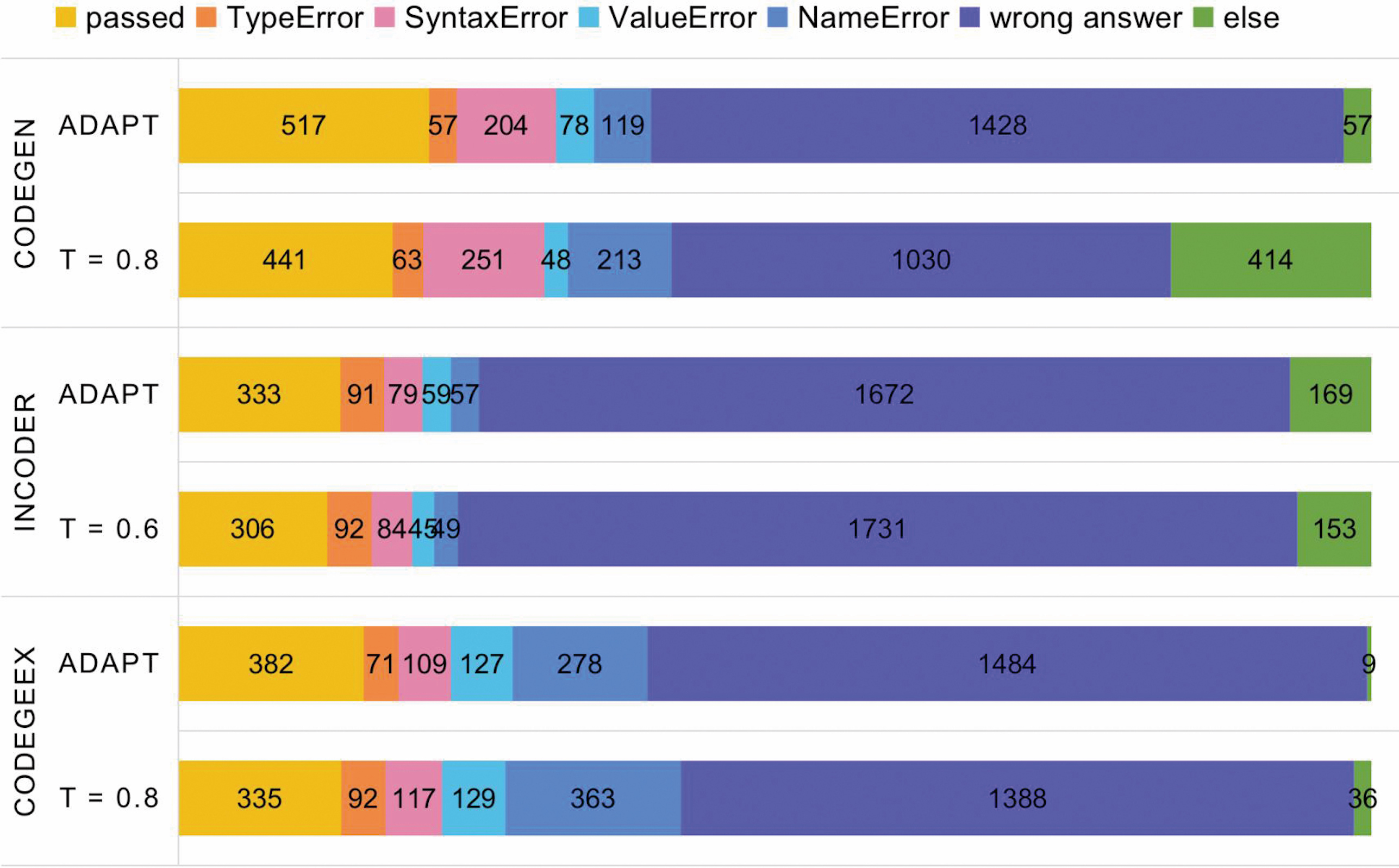}
    \caption{Evaluation of the generated code quality of AdapT sampling. Compared with the best-performed SP, AdapT sampling can reduce syntax errors and increase correct code sample numbers.}
    \label{error}
\end{figure}

\subsubsection{Code Quality Evaluation}
In this section, we analyze the quality of code generated by different sampling methods on the HumanEval dataset. We collect the execution results of the generated code of three models and present them in Figure \ref{error}. 
AdapT sampling can generate more correct samples (passed) than baselines on all models. 
Using AdapT sampling, CodeGen-2B, InCoder-6B, and CodeGeeX-13B can generate 517, 333, 382 correct codes, which outperforms SP up to 17.2\%,  8.8\%, 14.0\% higher than sampling with the SP method, respectively.

There are fewer TypeError and SyntaxError in the code generated by the three models using AdapT sampling than SP.  
By using AdapT sampling in CodeGen and CodeGeeX models, the occurrence of NameError can be reduced by 44.3\% and 23.4\%. In CodeGen and CodeGeeX, AdapT sampling reduces the incidence of NameError by 44.3\% and 23.4\%. The else section in Figure \ref{error} contains IndentationErrors, RecursionErrors, UnboundLocalErrors, RuntimeError, etc. These types of errors rarely occur in our AdapT sampling. The reason for this is that AdapT sampling uses a smaller $T$ inside the code block, which improves the coherence of the sampled code, and therefore reduces syntax errors.
For AdapT sampling, the most common error type is the wrong answer, showing that our method suffers from incorrect code logic. Taking steps to solve this issue can be a great improvement in the future. 
We present several case studies in the appendix.
\section{Future Work}

In this section, we discuss the remaining challenges of decoding strategies in code generation and provide some possible directions to facilitate other researchers.

\begin{itemize}

\item 
We recognize some challenging tokens within statements, but their distributions do not show a clear statistical pattern. In the method designing process, we have experimented with various temperature tuning functions, such as linear decay function, exponential decay function, etc., without obtaining substantial improvement.
In future work, we plan to use learning-based methods to adjust the temperature coefficients.

\item In practice, software development often relies on specific domain knowledge, such as private code libraries and code specifications. Existing decoding strategies ignore these issues. In the future, we can introduce domain knowledge into the decoding process, improving the usability of code-generation LLMs in real-world scenarios. 

\item As shown in Figure \ref{error},
LLMs are suffering from incorrect code logic, and generating code from scratch is very challenging. 
In the future, we can design a multi-stage decoding strategy, which steers LLMs to generate code progressively. For example, LLMs first generate a natural language plan and then generate an executable program based on the plan.

\end{itemize}

\section{Conclusion}
This paper is the first attempt at the LLM's decoding strategy for code generation.
We statistically analyze the loss distribution of source code and find out that code tokens can be categorized into challenging tokens and confident tokens. Moreover, challenging tokens often appear in the initial of a code block. Based on the insights, we propose AdapT sampling which dynamically adjusts the temperature coefficient through sampling and has proven its effectiveness on code generation datasets. 
Finally, we present several challenges and insights in developing a more advanced decoding strategy for code generation and we look forward to further exploring its potential in future research. 
\section{Acknowledgements}

This work is supported by the National Natural Science Foundation of China under Grant Nos.62192731, 62072007, 62192733, 61832009, 62192730, and 62332012, the National Key R\&D Program under Grant No.2023YFB4503801, and the Key Program of Hubei under Grant JD2023008.

\bibliography{ref}
\clearpage
\appendix
\section*{Appendix}

\subsection*{Hyperparameter Settings}
\begin{table}[htbp]
\centering
\caption{Hyperparameters used in Main Result Section}

\begin{tabular}{|c|c|c|c|c|}
\hline
Dataset & \multicolumn{2}{c|}{HumanEval} & \multicolumn{2}{c|}{MBPP}\\
\hline
 Model& $a$&$b$ & $a$&$b$\\
\hline
CodeGen & 0.8 & 0.4 & 0.7 & 0.6\\
\hline
InCoder & 0.8 & 0.2 & 0.7 & 0.4\\
\hline
CodeGeeX & 0.8 & 0.6 & 0.6 & 0.5\\
\hline
\end{tabular}
\label{hyper_table}
\end{table}

\begin{table}[htbp]
\centering
\caption{Hyperparameters used in Pass@1 Section}

\begin{tabular}{|c|c|c|c|c|}
\hline
Dataset & \multicolumn{2}{c|}{HumanEval} & \multicolumn{2}{c|}{MBPP}\\
\hline
 Model& $a$&$b$ & $a$&$b$\\
\hline
CodeGen & 0.2 & 0.01 & 0.2 & 0.1\\
\hline
InCoder & 0.3 & 0.01 & 0.2 & 0.01\\
\hline
CodeGeeX & 0.1 & 0.01 & 0.2 & 0.01\\
\hline
\end{tabular}
\label{hyper_table_pass1}
\end{table}

\subsection*{Analysis of challenging tokens}
\begin{figure}[htbp]
    \centering
    \includegraphics[width = 7.5cm]{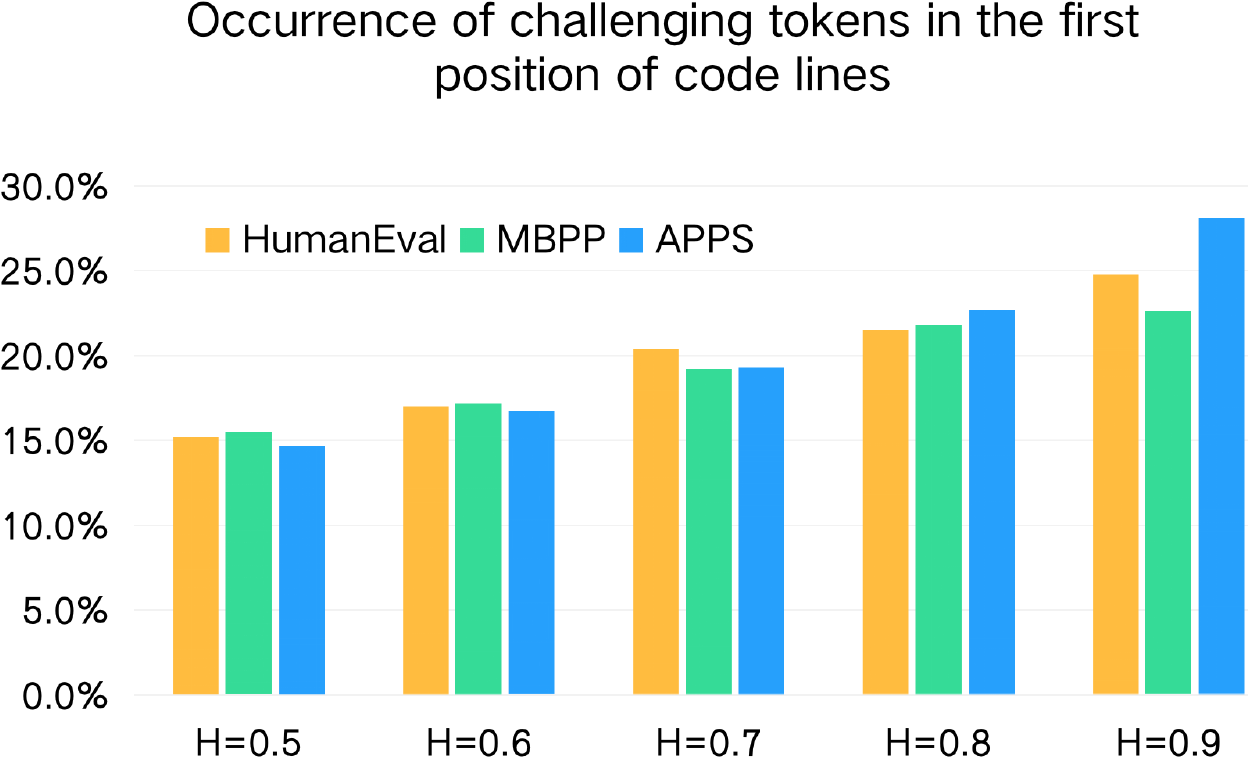}
    \caption{Occurence of challenging tokens in the first position of code lines}
    \label{challenging_token}
\end{figure}

We change $H$ from 0.5 to 0.9 and observe the occurrence of challenging tokens appearing in the first position of code lines, and the results of three datasets ($\ie$ HumanEval, MBPP, APPS) are shown in Figure \ref{challenging_token}.

The results show that as the $H$ increases, the chances of large loss tokens appearing at the beginning of the code line increase. This indicates that challenging tokens are more likely to appear at the beginning of the code lines.

\subsection*{Comparison with different $T$}
For brevity, we skip the data already shown in Table \ref{HumanEval_result} and Table \ref{mbpp_result} ($\ie$ $T$ = 0.2, 0.4, 0.6, 0.8).
Since greedy search can only generate one result, all the results of the greedy search are its pass@1 result in the following figures.

\begin{figure}[htbp]
    \centering
    \includegraphics[width = 7.5cm]{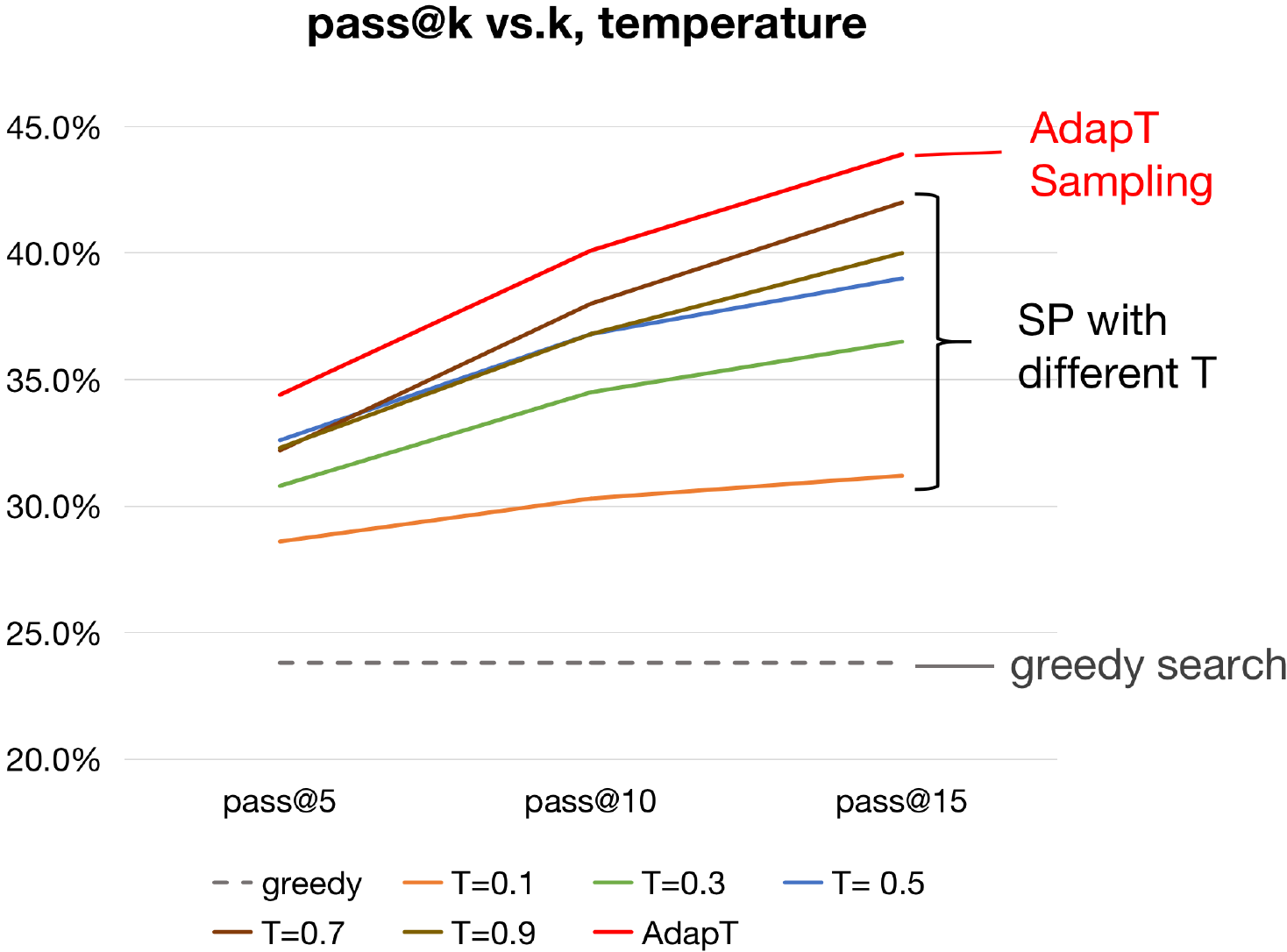}
    \caption{The performance of CodeGen-2B on the HumanEval with different $T$.}
    \label{codegen_he}
\end{figure}

\begin{figure}[htbp]
    \centering
    \includegraphics[width = 7.5cm]{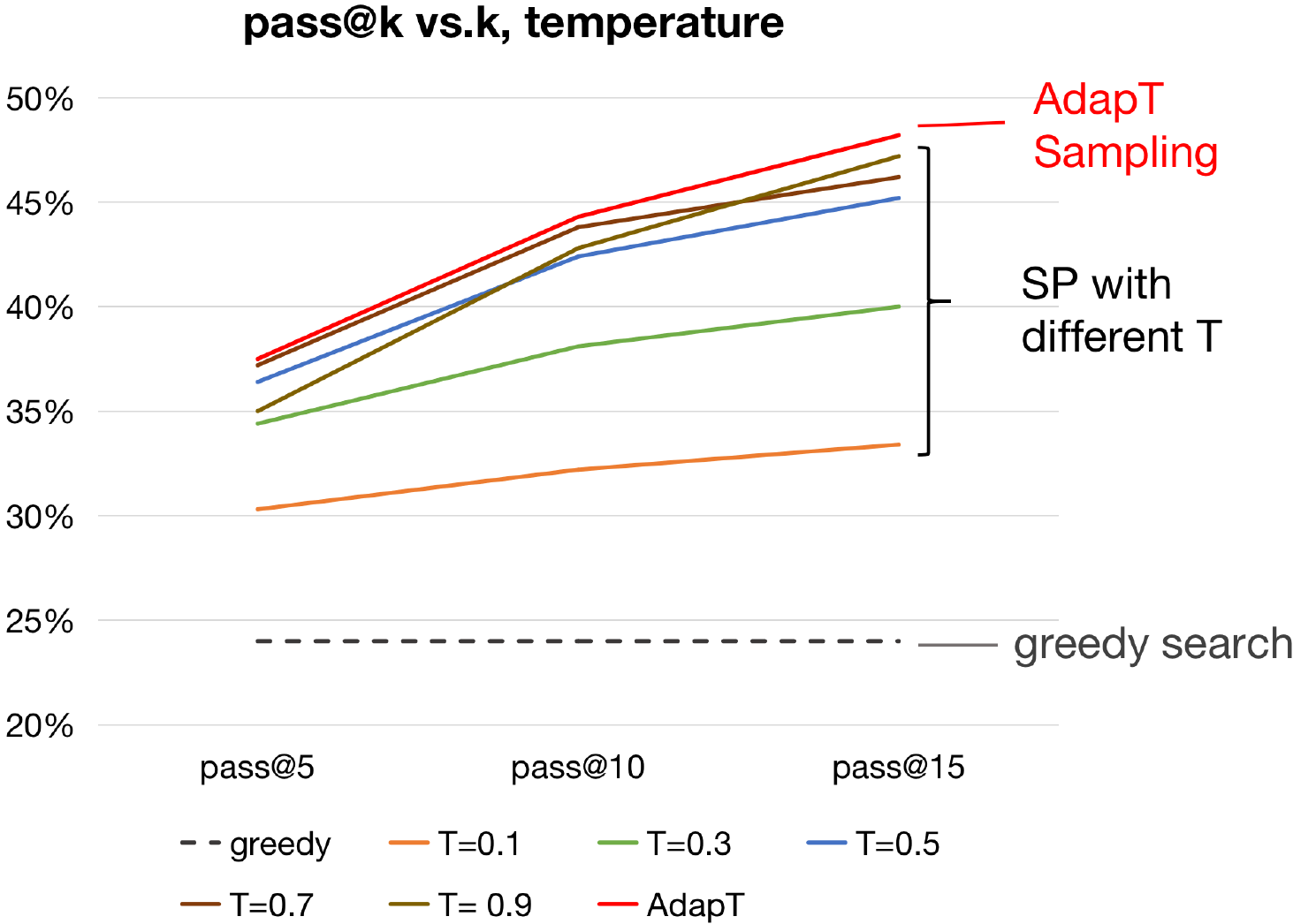}
    \caption{The performance of CodeGen-2B on the MBPP with different $T$.}
    \label{codegen_mbpp}
\end{figure}

\begin{figure}[htbp]
    \centering
    \includegraphics[width = 7.5cm]{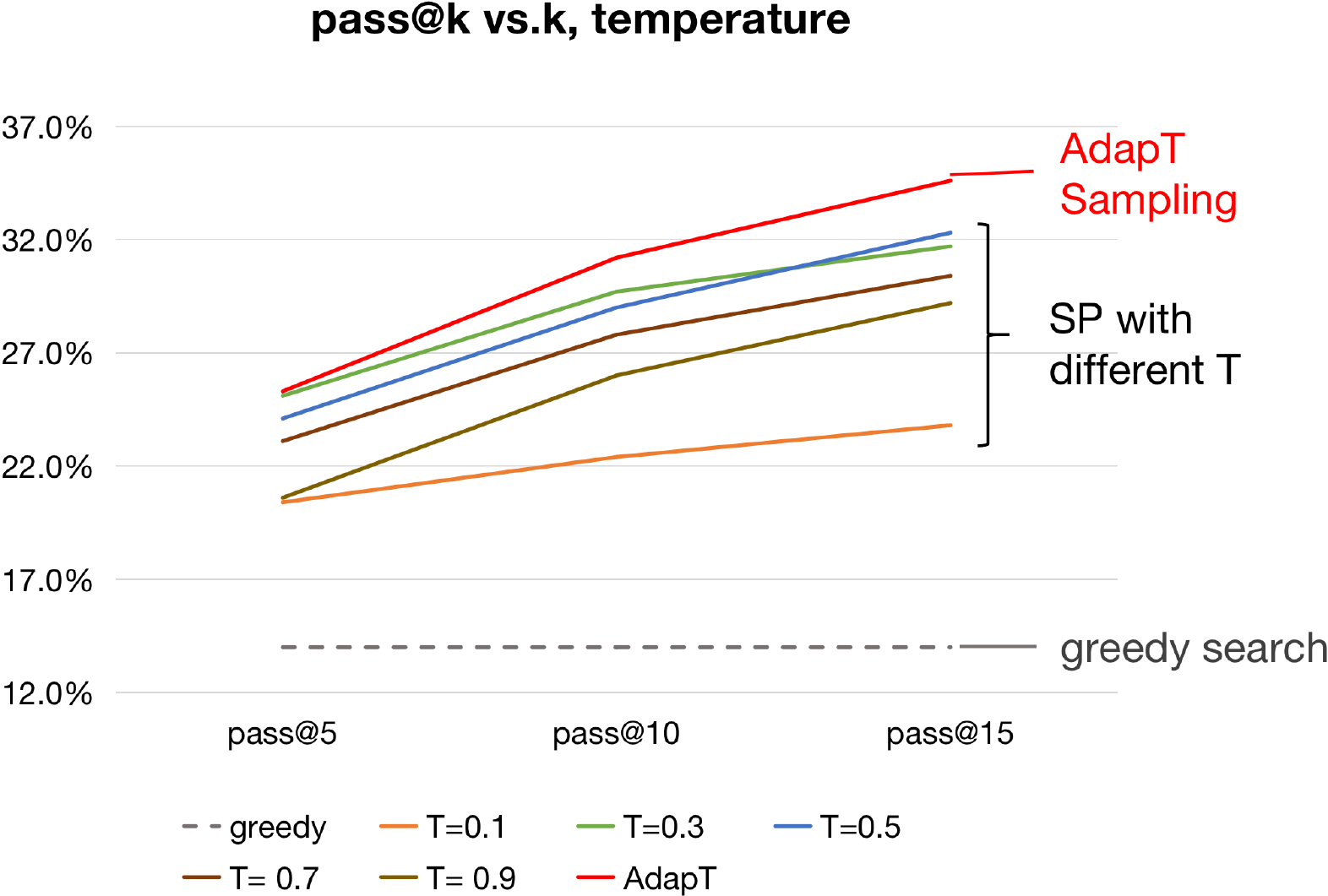}
    \caption{The performance of InCoder-6B on the HumanEval with different $T$.}
    \label{incoder_he}
\end{figure}

\begin{figure}[htbp]
    \centering
    \includegraphics[width = 7.5cm]{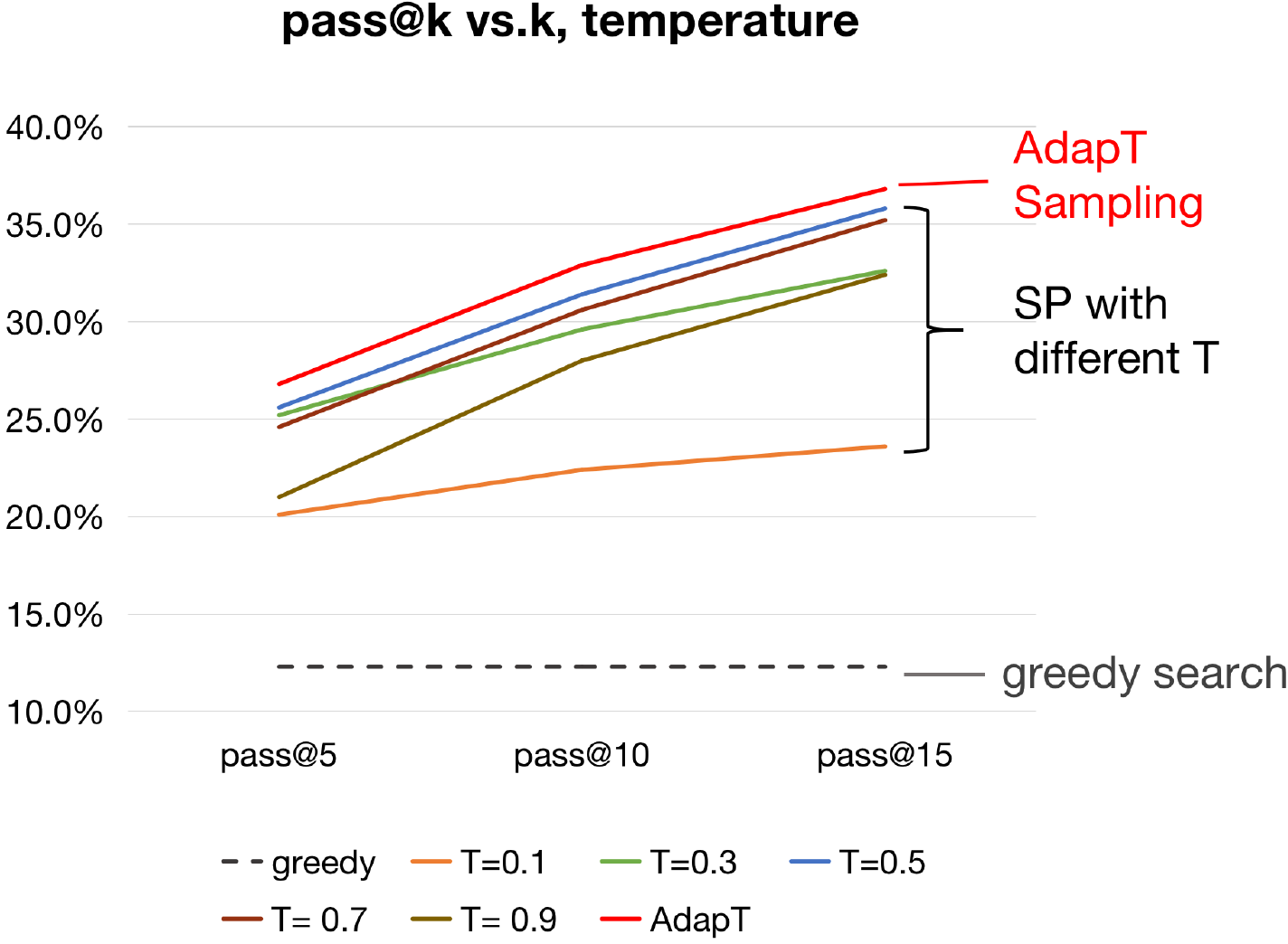}
    \caption{The performance of InCoder-6B on the MBPP with different $T$.}
    \label{incoder_mbpp}
\end{figure}

Figure \ref{codegeex_mbpp} shows the performance of CodeGeeX-13B on the MBPP with different $T$. 
Figure \ref{codegen_he} and \ref{codegen_mbpp} show the performance of CodeGen-2B on the HumanEval and MBPP datasets with different $T$.
Figure \ref{incoder_he} and \ref{incoder_mbpp} show the performance of InCoder-6B on the HumanEval and MBPP datasets with different $T$.

The results show that AdapT sampling outperforms the SOTA method in all $T$ settings on all models. This indicates that AdapT sampling can exceed the upper bound of the SOTA method.

\begin{figure}[htbp]
    \centering
    \includegraphics[width = 7.5cm]{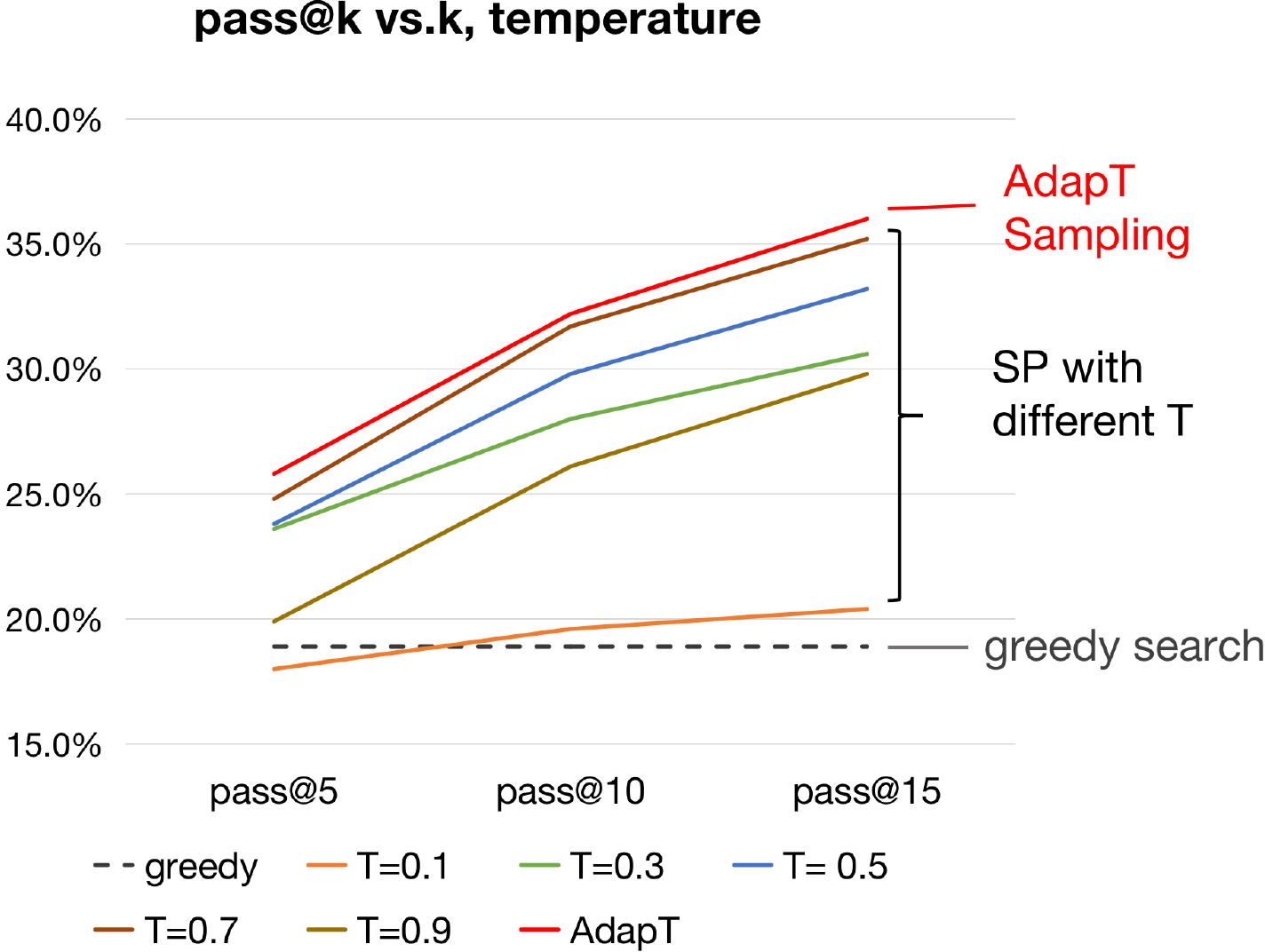}
    \caption{The performance of CodeGeeX-13B on the MBPP with different $T$.}
    \label{codegeex_mbpp}
\end{figure} 

\begin{figure*}[htbp]
    \centering
    \includegraphics[width = 13cm]{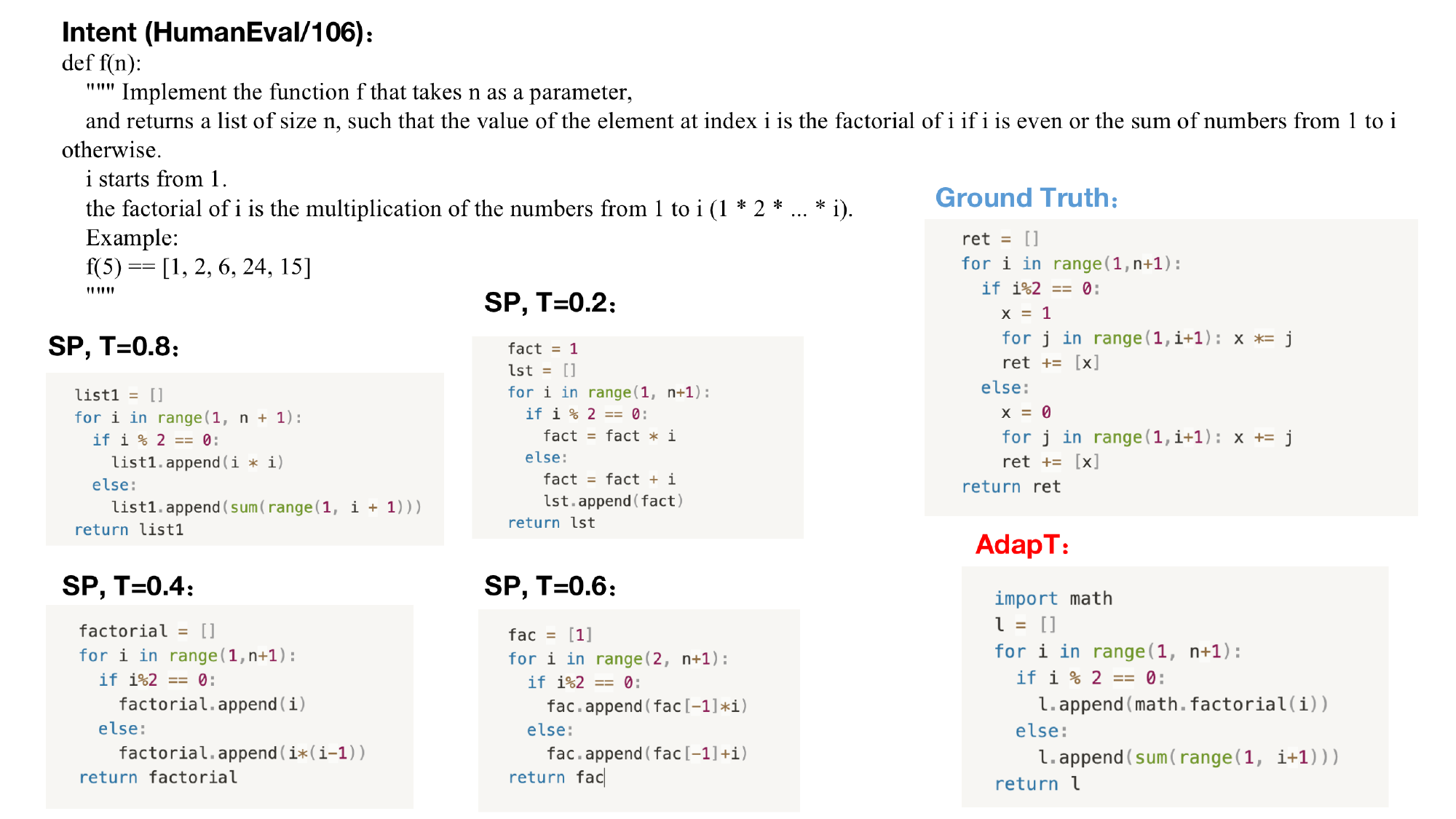}
    \caption{An example of generating code with AdapT sampling and SOTA method. Only AdapT sampling generates the correct answer.}
    \label{case_study}
\end{figure*} 


\subsection*{Case Study}
We show an example of code generation with AdapT sampling and SOTA baseline in Figure \ref{case_study}. It can be seen that the SOTA method fails to generate the correct answer under different $T$ settings, while AdapT sampling generates the code that passes all the test cases and with the correct code logic.

\end{document}